\documentclass[conference]{IEEEtran}
\usepackage[utf8]{inputenc}
\usepackage{graphicx}
\usepackage{caption}
\usepackage{amsmath}
\usepackage{float}
\ifCLASSINFOpdf
\else
\fi
\usepackage[utf8]{inputenc}
\usepackage{graphicx}
\usepackage{xcolor}
\usepackage{listings}
\usepackage{lipsum}
\begin{document}
%
\title{Device to Device Pairs Sharding based on Distance}

\makeatletter
\newcommand{\linebreakand}{%
  \end{@IEEEauthorhalign}
  \hfill\mbox{}\par
  \mbox{}\hfill\begin{@IEEEauthorhalign}
}
\makeatother

\author{\IEEEauthorblockN{K Prajwal}
\IEEEauthorblockA{Information Technology\\
National Institute of Technology Karnataka\\
Surathkal, India 575025\\
Email:kprajwal.191it222@nitk.edu.in}
\and
\IEEEauthorblockN{Tharun K}
\IEEEauthorblockA{Information Technology\\
National Institute of Technology Karnataka\\
Surathkal, India 575025\\
Email:tharunk.191it255@nitk.edu.in}
\linebreakand
\IEEEauthorblockN{Navaneeth P }
\IEEEauthorblockA{Information Technology\\
National Institute of Technology Karnataka\\
Surathkal, India 575025\\
Email: navaneethp.191it132@nitk.edu.in}
\and
\IEEEauthorblockN{Ishwar Chandra Mandal}
\IEEEauthorblockA{Information Technology\\
National Institute of Technology Karnataka\\
Surathkal, India 575025\\
Email:ishwarmandal.191it122@nitk.edu.in}
\linebreakand
\IEEEauthorblockN{Kiran M}
\IEEEauthorblockA{Information Technology\\
National Institute of Technology Karnataka\\
Surathkal, India 575025\\
Email:kiranmanjappa@nitk.edu.in}
}

%


\maketitle

\begin{abstract}
In the conventional cellular system, devices are not allowed to communicate directly with each other in the licensed cellular bandwidth and all communications take place through the base stations. The users requirements has led the technology to become fast and faster. Multimedia rich data exchange, fast service, high quality voice calls, newer and more demanding applications, information at fingertips, everything requires technology and communication between devices. A constant need to increase network capacity
for meeting the users growing demands has led to
the growth of cellular communication networks from the first
generation(1G) to the fifth generation(5G). There will be crores
of connected devices in the coming future . A large number of
connections are going to be heterogeneous,
demanding lesser delays, higher data rates, superior throughput and enhanced system
capacity. The available spectrum
resources are limited and has to be flexibly used by mobile
network operators to cope with the rising demands. An
emerging facilitator of the upcoming high data rate demanding
next-generation networks are device-to-device(D2D) communication. This paper has developed a model that establishes Device-to-Device (D2D) communication between two end-users without involving the eNB (evolved Node B). We have sharded the UEs and CUs based on the criteria of DISTANCE. To do so, we used the K-means clustering method. \\

Keywords : Cellular communication networks, D2D communication, Distance,  K-means clustering method.
\end{abstract}


%
\IEEEpeerreviewmaketitle

\section{Introduction}
Nowadays, the number of hand-held devices is increasing drastically, with rising demand for higher data rate
applications. Analysts predict an explosive growth in traffic demand on mobile broadband systems over the coming years due to the popularity of streaming video, gaming, and other social media services. In order to meet the needs of the next
generation applications, the present data rates need a
refinement. The fifth generation(5G) networks are expected
and will fulfill these increasing demands. A competent
technology of the NGNs(next generation networks) is Device-to-Device (D2D) Communication also called as Sidelink Communication, which is expected to play
an indispensable role in the approaching era of wireless
communication. The use of D2D communication did not gain
much importance in the previous generations of wireless
communication. A key motivation for D2D connectivity is its potential for operators to offload traffic from the core network, and the framework for a new communication paradigm to support social networking through localization. In 5G networks, it is expected to be a vital
part. The rising trends pave way for this emerging
technology. With the introduction of device-to-device (D2D)
communication, direct transmission between devices is
possible. This is expected to improve the reliability of the link between the devices, enhance spectral efficiency and system
capacity, with reduced latency within the networks. Such a
technique is essential for fulfilling the chief goals of the
mobile network operators (MNOs).\\

\begin{figure}[htp]
    \centering
    \includegraphics[width=8cm]{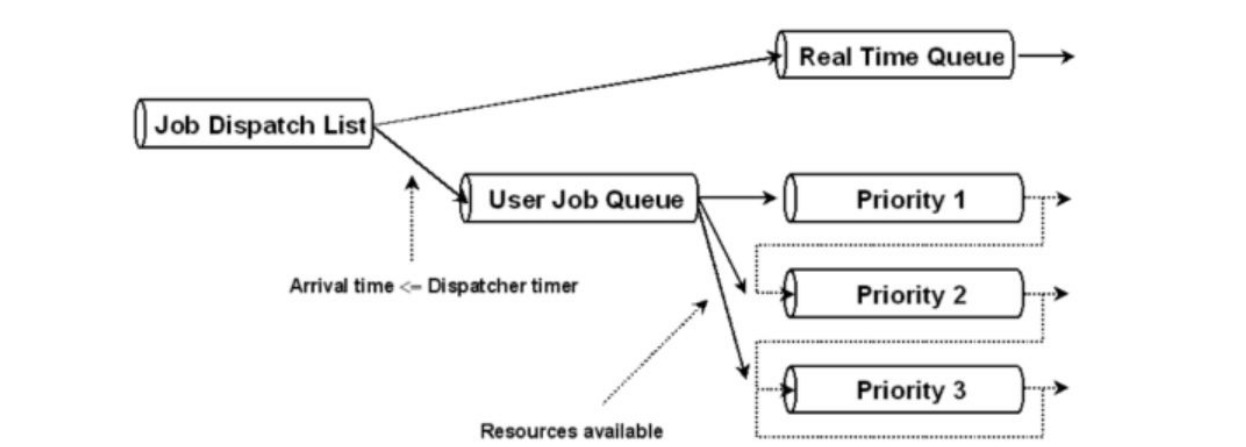}
    \caption{D2D Features}
    \label{fig:flowchart }
\end{figure}

D2D communication allows communication between two
devices, without involving the Base Station (BS), or
the evolved NodeB (eNB). ProSe enabled UE, works with close-proximity UE devices. Thus, ProSe enables direct communication between nearby devices without involving the
base station. Thus, it minimizes the delay. Here, as the communication link is not established through the eNB, it just supports the communication, but will not be part of it. Proximate devices can directly
communicate with each other by establishing direct links. Due
to the small distance(of about 500m or less) between the D2D users, it supports power
saving within the network, which is not possible in the case of
conventional cellular communication. It promises
improvement in energy efficiency, throughput and delay. It
has the potential to effectively offload traffic from the network
core. It provides ultra-low Latency communication. Hence, it is a very flexible technique of communication
within the cellular networks.

 
\section{Literature Survey}

Current work undergoing under D2D Communication:\\

Device-to-device communication is widely being accepted
by the mobile stakeholders and they believe it to be a big
success in wireless technology. Qualcomm, LTE-A and IEEE
802.15.4g (SUN) are at present involved in the standardization
activities of D2D communication over the licensed band.
\\\\
IEEEE 802.15.4g was first released as an amendment to the
low rate WPAN in April 2012. It supports three different
modulation techniques, FSK, DSSS and OFDM. Maximum
data rate supported is upto 200kbps with a maximum range of
about 2-3 km. SUN is highly energy efficient, which is a very
attracting feature. Other IEEE standards include IEEE
802.15.8 (for PHY/MAC specification of D2D) and IEEE
802.16n.
\\\\
D2D communications with and without infrastructure are
being studied by 3GPP. Proximity-based Services (ProSe) and
Group Communication System Enablers for LTE
(GCSE-LTE) are discussed in [4], [5] and [6]. D2D ProSe
considers various aspects of D2D communication, including
one-to-one, one-to-many and one-hop relay and also addresses
switching of mode between D2D mode and cellular mode.
ProSe includes working on identifying UEs in proximity (peer
discovery, and establishing direct links between them, so as to
enable communication between them, either directly or through
a locally routed path via the eNB.\\

After reading through the above sources, we have concluded to use the k-means clustering method. This sharding method is relatively simpler to implement compared to other methods like Quality Threshhold Clustering, Expectation Maximization Clustering, Mean shift. Also, once clusters and their associated centroids are identified, it is easy to assign new objects to a cluster based on the object's distance from the closest centroid.

\section{Problem Statement}

Given D2D and CU, divide the UE's into shards so that each of the shards contains at least one CU. You can separate the user equipment based on criteria like distance or SINR, or resources available. Divide the D2D user equipment into pairs and calculate the SINR  of all the pairs.
\begin{figure}[htp]
    \centering
    \includegraphics[width=8cm]{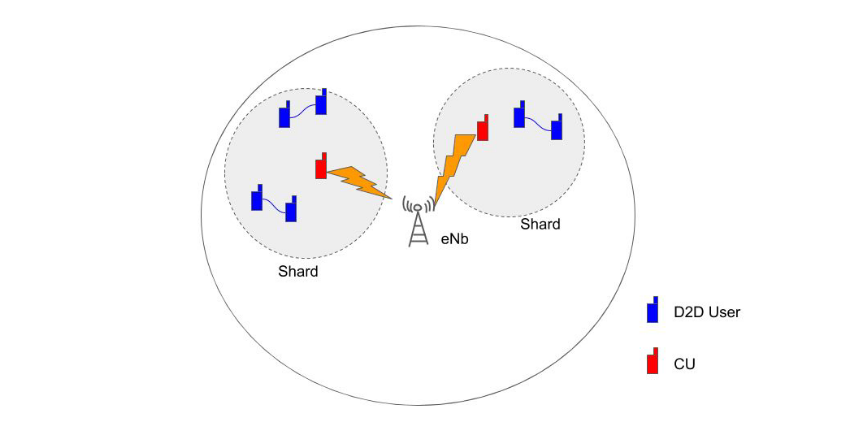}
    \caption{Flowchart of Model }
    \label{fig:flowchart}
\end{figure}

\subsection{Objectives}

\begin{enumerate}
\item  Dividing the devices based on distance using the K-means algorithm.
\item  Dividing the D2D devices into pairs. 
\item  Calculating the SINR of the each D2D pairs. .
\end{enumerate}

\section{Methodology}

\subsection{\textbf{D2D and CU nodes Creation}}
\begin{lstlisting}
 NodeContainer enbNode;
    enbNode.Create(1);

    NodeContainer ueNodes;
    NodeContainer ueClients;
    NodeContainer ueServers;
    std::vector<NodeContainer> uePairs(n);

    for (uint32_t i = 0; i < n; ++i) {
        NodeContainer pair;
        pair.Create(2);

        ueClients.Add(pair.Get(0));
        ueServers.Add(pair.Get(1));

        ueNodes.Add(pair);

        uePairs[i].Add(pair);
    }

    NodeContainer cuNodes;
    cuNodes.Create(k);

  \end{lstlisting}
  Here we created 2n D2D nodes, i.e n D2D pairs and k CU nodes and one eNb Node. The difference between the D2D nodes and the cellular user nodes is that the sidelink is enabled in the D2D nodes. This means that the D2D nodes can communicate with each other using the sidelink, whereas the cellular user nodes communicate with each other with the help of eNb using the uplink and the downlink. Therefore, when we create the cellular nodes, we do not enable sidelink.
  
\subsection{\textbf{Position Creation of D2D nodes and CU nodes}}
We have to give the position for the D2D nods and the CU nodes, and this is done by using a loop and placing them at random position using the rand function. We have used points as a vector that takes these values and is passed on the K-means clustering function. 
\begin{lstlisting}
//Position of the nodes
    Ptr<ListPositionAllocator> positionAllocEnb =
    CreateObject<ListPositionAllocator>();
    positionAllocEnb->Add(Vector(0.0, 0.0, 0.0));

    random_device seed;
    mt19937 gen(seed());
    uniform_real_distribution<float> dist(0.0, 50.0);

    int x1, y1, x2, y2;
    vector<Point> points;

    Ptr<ListPositionAllocator> positionAllocUe[n];
    for (int i = 0; i < n; i++) {
        x1 = dist(gen);
        y1 = dist(gen);

        x2 = x1 + 10;
        y2 = y1 + 10;

        points.push_back(Point((x1 + x2) / 2, 
        (y1 + y2) / 2, 0,
        uePairs[i].Get(0)->GetId(),
        uePairs[i].Get(1)->GetId()));

        positionAllocUe[i] = 
        CreateObject<ListPositionAllocator>();
        positionAllocUe[i]->Add(Vector(x1, y1, 0.0));
        positionAllocUe[i]->Add(Vector(x2, y2, 0.0));
    }

    Ptr<ListPositionAllocator> positionAllocCu[k];
    for (int i = 0; i < k; i++) {
        x1 = dist(gen);
        y1 = dist(gen);
        points.push_back(Point(x1, y1, 1, 
        cuNodes.Get(i)->GetId()));
        positionAllocCu[i] = 
        CreateObject<ListPositionAllocator>();
        positionAllocCu[i]->Add(Vector(x1, y1, 0.0));
    }
  \end{lstlisting}
The above is an algorithm that we used, and it is used to give a random position for the D2D and CU devices. 

\subsection{\textbf{The point structure}}
We have a structure known as the point structure that contains the following variables : 
\begin{enumerate}
\item double x, y: the variables x and y tells us the x and y coordinates of the nodes. 
\item int cluster: this variable tells us which cluster the node is in. 
\item int node ID: this tells us the ID of the node. 
\item int node type : this gives the information about the type of the node.
\item IPv4Address node IP: this variable gives us the Ipv4 address of the node.
\item double minDist : this variable gives us the minimum distance. 
\end{enumerate}

\begin{lstlisting}
struct Point {
    double x, y;
    int cluster;
    uint32_t nodeId1;
    uint32_t nodeId2;
    int nodeType;
    Ipv4Address nodeIp1;
    Ipv4Address nodeIp2;
    double minDist;

    Point(double x, double y, int type, 
    	uint32_t node1, uint32_t node2 = -1)
        : x(x)
        , y(y)
        , cluster(-1)
        , nodeId1(node1)
        , nodeId2(node2)
        , nodeType(type)
        , minDist(__DBL_MAX__)
    {
    }

    double distance(Point p)
    {
        return (p.x - x) * (p.x - x) + 
        	(p.y - y) * (p.y - y);
    }
};
  \end{lstlisting}

\subsection{\textbf{K-Means Clustering }}

K-means is a technique that was initially introduced in Signal processing that aims to divide/separate n elements into k clusters. Each element belongs to a cluster with the nearest mean. In the algorithm, we consider the distance between the elements, and K-means clustering minimizes the within-cluster Euclidean distances.
\\\\
\textbf{Distance} =  $((x_1 - x_2 ) ^2 +(y_1 - y_2) ^2)^{0.5}$
\\\\
The below is a code snippet for the K-means clustering algorithm.
\begin{lstlisting}
vector<Point> centroids;
    srand(time(0));
    for (int i = 0; i < k; ++i) {
    for (vector<Point>::iterator x = points->begin(); 
               x != points->end(); ++x) {
            if (x->node_type == 1 &&
                  x->cluster == -1) {
                centroids.push_back(*x);
            }
        }
  for (vector<Point>::iterator c = begin(centroids);
           c != end(centroids); ++c) {
  int clusterId = c - begin(centroids);
  for (vector<Point>::iterator it = points->begin(); 
    it != points->end(); ++it) {
                Point p = *it;
                double dist = c->distance(p);
                if (dist < p.minDist) {
                    p.minDist = dist;
                    p.cluster = clusterId;
                }
                *it = p;
            }
        }
    }
    vector<int> nPoints;
    vector<double> sumX, sumY;

    for (int j = 0; j < k; ++j) {
        nPoints.push_back(0);
        sumX.push_back(0.0);
        sumY.push_back(0.0);
    }

   for (vector<Point>::iterator it = points->begin();
        it != points->end(); ++it) {
        int clusterId = it->cluster;
        nPoints[clusterId] += 1;
        sumX[clusterId] += it->x;
        sumY[clusterId] += it->y;

        it->minDist = __DBL_MAX__;
    }

    for (vector<Point>::iterator c = begin(centroids)
            ; c != end(centroids); ++c) {
        int clusterId = c - begin(centroids);
        c->x = sumX[clusterId] / nPoints[clusterId];
        c->y = sumY[clusterId] / nPoints[clusterId];
    }
  \end{lstlisting}
The problem is computationally challenging; that is, it is an NP-hard problem; however, we can converge it into a local optimum using efficient heuristic algorithms. Our project divided the UE's into 3 clusters based on the distance, and we made sure that there is at least one cellular user in each node. We created a centroid vector and used this vector to create shards. 
\begin{figure}[H]
    \centering
    \includegraphics[width=8cm]{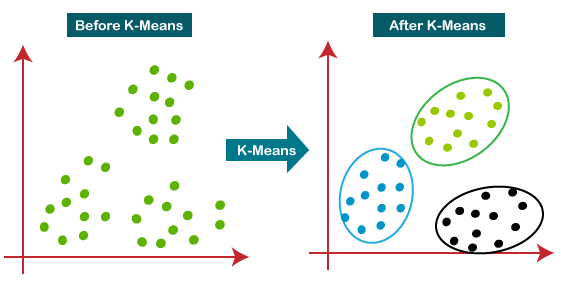}
    \caption{Figure depicting the K-Means Clustering algorithm }
    \label{fig:flowchart}
\end{figure}
\subsection{\textbf{Calculating SINR }}
 In telecommunication engineering and information theory, the signal-to-interference-plus-noise ratio (SINR) (also known as the signal-to-noise-plus-interference ratio (SNIR)) is a quantity that gives the theoretical values of the upper bounds on channel capacity in wireless communication systems such as networks. It is provided by the formula :
 \\ 

\[
  SINR  =\frac{power-of-a- specific- signal- of-interest}{ the- sum- of -the -interference -power}
\]

We found out that there is a call back function that takes in multiple parameters, including SINR. We used the config::connect and gave it a path of /NodeList//DeviceList//\$ns3::LteUeNetDevice/ComponentCarrierMapUe/*/\\LteUePhy/ReportCurrentCellRsrpSinr to calculate the SINR. Now we used the SINR parameters from this 
\begin{figure}[H]
    \centering
    \includegraphics[width=8cm]{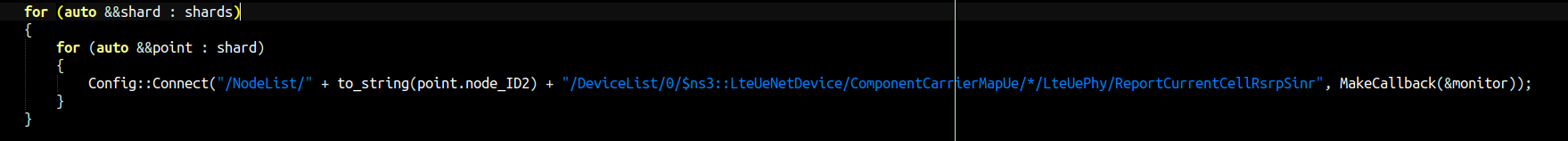}
    \caption{Code Snippet for declaring the path that calculates the SINR }
    \label{fig:flowchart}
\end{figure}

\section{Results and Analysis}

We were able to shard the n D2D pairs, and in each of the shard we were able to include a cellular user node. The node type has two values, 0 and 1, 0 = D2D user and 1 = Cellular user. We can see that the nodes with node id 24, 25 and 26 have node type 1 and each of them are in a different shard. Since there are three shards and we have three cellular user with three different shard numbers we can conclude that there is at least one cellular user in each shards. Also we can see that 6 D2D users are there in shard  1 implying 3 D2D pairs and then there are  4 D2D users in shard 2, implying that there are 2  pairs of D2D users and then there are 10 D2D users in shard 0, implying 5 pair of D2D user. 
\begin{figure}[htp]
    \centering
    \includegraphics[width=9cm]{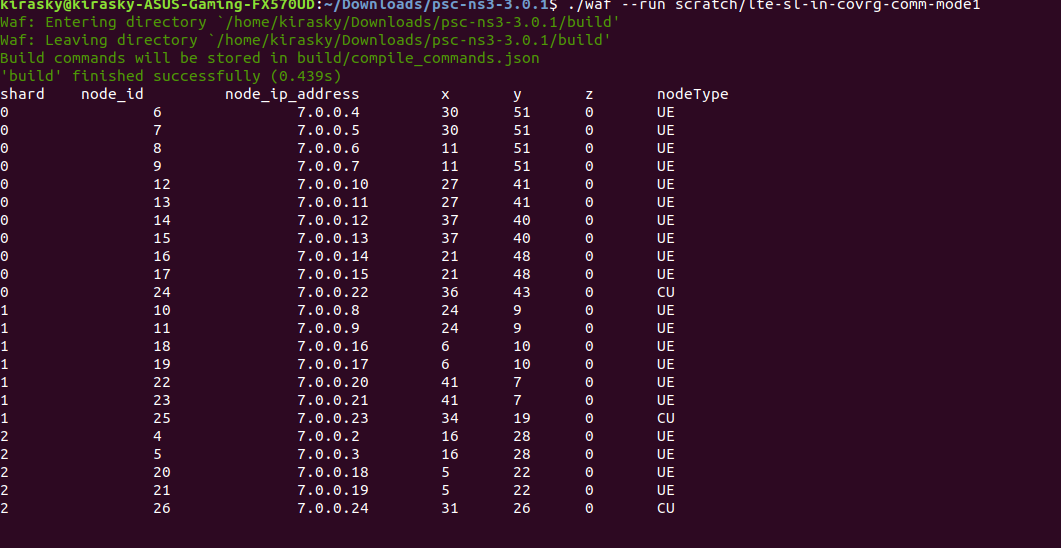}
    \caption{Screenshot of the Sharding of N D2D pairs }
    \label{fig:flowchart}
\end{figure}

\begin{figure}[H]
    \centering
    \includegraphics[width=5cm]{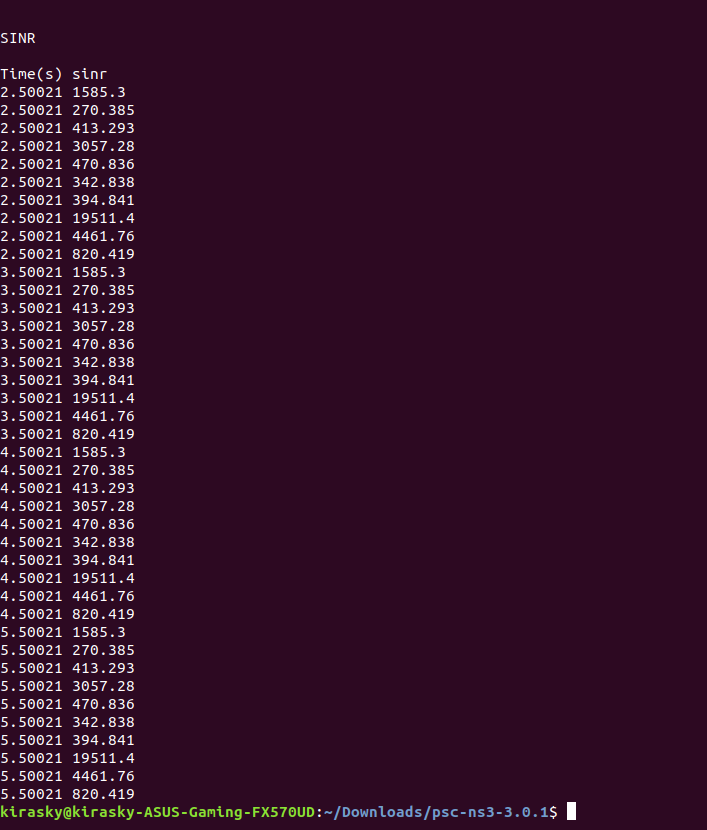}
    \caption{Screenshot of the SINR calculation at regular intervals }
    \label{fig:flowchart}
\end{figure}
The SINR is outputted at regular intervals starting from 2.5 seconds till 6 seconds, which is the simulation's end time. The SINR is displayed at a regular interval of 1 second. There are many SINR's because there are many packets, and the SINR is calculated every time a packet is received.

\section{Conclusion}
We were able to make a shard in which each shard had one cellular user and some D2D users. We created the shard based on distance using K-means clustering algorithm and printed out the IP address of each devices and their shard number.

\section*{Acknowledgment}

Principally, we are thankful to the National Institute of Technology Karnataka, Information Technology Department and Prof. Kiran M, Professor, Information Technology Department, National Institute of Technology Karnataka, for this wonderful opportunity.




%

\end{document}